\documentclass[a4paper,twocolumn,10pt]{revtex4}
\usepackage{amsfonts}
\usepackage{color}
\usepackage{eepic}
\usepackage{epic}
\usepackage{ulem}
\usepackage{graphicx}
\usepackage{texdraw}
\usepackage{epsfig}
\begin{document}
\title{Hot QCD equations of state and response functions for quark-gluon plasma}
\author{Vinod Chandra$^1$}
\email{vinodc@iitk.ac.in}
\author{Akhilesh Ranjan$^1$}
\email{akranjan@iitk.ac.in}
\author{V. Ravishankar$^{1,2}$}
\email{vravi@iitk.ac.in,vravi@rri.res.in}
\affiliation{$^1$Department of Physics, Indian Institute of Technology Kanpur,
 UP, India, 208 016}
\affiliation{$^2$Raman Research Institute, C V Raman Avenue, Sadashivanagar, Bangalore, 560 080, India}
\date{\today}
\begin{abstract}
We study the response functions (chromo-electric susceptibilities)
of  quark-gluon plasma as a function of temperature in the presence of 
interactions. We consider two equations of state for hot QCD. The first one is  fully perturbative, 
 of  $O(g^5)$ EOS and, and the second one which is  $O[g^6\ln(1/g)+\delta]$,   incorporates some non-perturbative effects.  Following a recent work (Physical Review {\bf C 76}, 054909(2007)), the interaction effects contained in the EOS are encapsulated  in terms of effective 
chemical potentials($\tilde\mu$) in the equilibrium distribution functions for the partons.By using them in another recent formulation of the response functions({\tt arXiv:0707.3697}),  we  
determine explicitly the chromo-electric 
 susceptibilities for QCD plasma. We find that it shows large deviations from the ideal behavior. We further study the modification in the heavy quark potential due to the medium effects. In particular, we determine the temperature dependence of the screening lengths by fixing the effective coupling constant $Q$ which appears in the transport equation by comparing the screening in the present formalism with exact lattice QCD results. Finally, we study the dissociation phenomena 
of heavy quarkonium states such as $c\bar{c}$ and $b\bar{b}$, and determine the dissociation temperatures.  Our results are in good agreement with  recent lattice results.

\vspace{2mm}
\end{abstract}
\maketitle

\section{Introduction}  
It is expected that at high temperatures ($T \sim 150-200 MeV$) and high densities ($\rho \sim 10 Gev/fm^{3}$) nuclear matter undergoes a deconfinement  transition to the quark-gluonic phase. 
This phase is under intense investigation in heavy ion collisions, and already, interesting results
have been reported by  Relativistic Heavy Ion Collider(RHIC) experiments \cite{expt}. As an important development, flow measurements\cite{star-report} suggest that close to the transition temperature $T_c$, the quark-gluon plasma (QGP) phase is strongly interacting ---
showing an almost perfect liquid behavior, with very low viscosity to entropy ratio --- rather than  showing a behavior close to that of an ideal gas. See Ref. \cite{new-matter} for a 
  comprehensive review of experimental observations from RHIC, and   Ref. \cite{expt,exp-status1,exp-status2,s-a-bass,csorgo} for other recent
experimental results. On the other hand,
lattice computations \cite{lattice,lattice-new} also suggest that QGP is strongly interacting even at $T=2T_c$. This finding has been reproduced by a number of other theoretical studies ---
 by employing AdS/CFT correspondence in the strongly interacting regime of QCD\cite{dtson},  by molecular dynamical simulations for classical strongly coupled systems\cite{shuryak}, and  by  model calculations with Au-Au data from RHIC \cite{bair,jyo}.  \footnote{In the backdrop of the above developments, 
a number of standard diagnostics, such as $J/\psi$ suppression and strangeness enhancement, which have been proposed to probe QGP also need to be re-examined. It is also of importance to address other transport properties, production and equilibration dynamics, and the physical manifestations of pre-equilibrium evolution.} 

If  this be the case, as it indeed appears to be, then the plasma interactions would be largely in the non-perturbative regime; in this regime,  few analytic techniques  are available for a robust theoretical analysis. Effective interaction approaches are needed.
 In  this direction, considerable work has already been done and we refer the reader to Ref. \cite{matsui,elze,vr1,vr2,akranjan,shur2,shur3,blaizot1} for some of the theoretical results.

The effective approaches emphasize the collective origin of the plasma properties which can be best
understood within a semi-classical framework. Indeed, in a recent work
\cite{fluid}, the  successes of  hydrodynamics in interpreting and understanding the experimental observations from RHIC has been reviewed.  Since more exciting and discerning data is expected from LHC experiments soon, and given the above context, it is worthwhile
exploring semi-classical techniques to understand the properties of QGP in heavy ion collisions.
In this context, it is known by
now \cite{pisarski,braaten-1,braaten-2,nair} that a classical behavior emerges 
naturally when one considers hard thermal loop(HTL) contributions. A local 
formulation of HTL effective action has been obtained by Blaizot and Iancu who
have succeeded in rewriting the HTL effective theory as a kinetic theory with
a Vlasov term \cite{blaizot,blaizot-2,blaizot-3,blaizot-4}. A significant development in this direction
is the realization that the HTL effects are, in fact, essentially classical and that they are much
easier to handle within the frame work of classical transport equations \cite{cm1,cm4,cristina}.  Thus,
the semi-classical techniques appear hold the promise of providing tools to understand the bulk properties of QGP.

The present paper continues the theme, and its
 central aim  is to combine the kinetic equation approach which yields the transport properties, with the hot QCD equations of state to make predictions which can be perhaps tested in heavy ion collisions. Recently, Ranjan and Ravishankar have developed a systematic approach to determine  fully the response functions  of QGP, with a special emphasis on the color charge as a dynamical variable \cite{akranjan}.
 In parallel,
Chandra, Kumar and Ravishankar have succeeded in adapting  two hot QCD EOS to make predictions for heavy ion collisions \cite{chandra1}. They have shown that  the 
interaction effects which modify  the equations 
of state can be expressed by 
 absorbing them into effective fugacities ($z_{q,g}$) of otherwise free or weakly interacting quasi quarks and gluons. 
 Since the analysis in Ref. (\cite{akranjan}) was illustrated only for (the academically interesting) case of ideal quarks and gluons, it is but natural to bring the two studies together and explore what the hot QCD EOS have to predict for heavy ion collisions.
We take up this program in this paper.

The main result of this paper is the determination of the modification that  the heavy quark potential undergoes in a medium constituted by interacting QGP, as predicted by the two EOS which we consider. After determining the screening length as a function of temperature, we focus on the Cornell potential\cite{cornell}  and study the dissociation mechanism for $c\bar{c}$ and $b\bar{b}$ states. 
The results are rather surprising and may as well signal the inapplicability of these EOS to describe
the deconfined phase. On the other hand, if the transition from the confined to the deconfined state
is not a phase transition as several studies predict \cite{phaseT}, it may still be possible to 
attribute some physical significance to the predictions of these EOS. We undertake the project here.
We show that, by using  one of the phenomenological
EOS is quite a good approximation to the more rigorous lattice results, the value of the phenomenological
coupling constant that occurs in the Boltzmann equation can be fixed.
Ultimately, the physical viability or otherwise of the results need to be established by comparing them
repeating the analysis of \cite{chandra1} with the lattice EOS. That will be taken up in a separate paper.

We consider two specific  hot QCD equations of state: The first, which we call EOS1 is
perturbative, with contributions up to 
$O(g^5)$\cite{arnold,zhai}. The second EOS has  a free parameter $\delta$, and is evaluated
upto $O[g^6\log(1/g)]$\cite{kaj1}. We denote it by EOS$\delta$.  $\delta$ may be
fine tuned to get a reasonably good agreement \cite{kaj1} with the lattice results
\cite{fkarsch}, which we exploit here.  Both the EOS are expected to be valid for $T >2T_c$ \cite{kaj1}, and  EOS$\delta$ is reliable beyond $T\sim 4T_c$.

The paper is organized as follows:  In section II, we  introduce  the two
hot QCD equations of state and outline the recently developed 
method\cite{chandra1} to adapt them for making definite predictions for QGP at RHIC 
and the forthcoming experiments at LHC. In section III, we  obtain the expressions for 
the response functions of interacting QGP and in section IV, we  study
their temperature dependence in detail. In section V, we  study the 
modifications in heavy quark potential due to the  hot QCD medium. We 
further study the temperature dependence of the Debye screening lengths in 
hot QCD.  We investigate the ``melting phenomena" of heavy quarkonia such 
as $J/\Psi$ and $b\bar{b}$ in the  medium, and extract the dissociation temperature. In doing so we also relate the phenomenological charge that occurs in the transport
equation to  lattice and experimental observables. We conclude the paper in section VI.

\section{Hot QCD equations of state and their quasi-particle description}
There are various equations of state proposed for QGP at RHIC. These include 
non-perturbative lattice EOS \cite{fkarsch}, hard thermal loop(HTL) resumed 
EOS\cite{htleos} and perturbative hot QCD equations of state
\cite{zhai,arnold,kaj1}. In the present paper, we seek to determine the 
chromo-electric response functions for QGP by employing two EOS:
(i) the fully perturbative $O(g^5)$ hot QCD EOS 
proposed by Arnold and Zhai\cite{arnold} and Zhai and Kastening
\cite{zhai}, and (ii) The EOS of   $O[g^6(\ln(1/g)+\delta)]$  determined by Kajantie {\it et al}
\cite{kaj1}, by incorporating contributions from non-perturbative 
scales, $gT$ and $g^2T$. We employ the method
recently formulated by Ranjan and 
Ravishankar\cite{akranjan} to extract the chromo-electric permittivities of the medium.  EOS1 reads
\begin{eqnarray}
\label{eqn1}
P_{g^5}&=&\frac{8\pi^2}{45\beta^4}\bigg \lbrace (1+\frac{21N_f}{32})-\frac{15}{4}(1+\frac{5N_f}{12})\frac{\alpha_s}{\pi}
+30(1+\frac{N_f}{6})(\frac{\alpha_s}{\pi})^{\frac{3}{2}} \nonumber\\
&&
+\bigg[(237.2+15.97N_f-0.413 N_f^2 +\frac{135}{2}(1+\frac{N_f}{6})\ln(\frac{\alpha_s}{\pi}\nonumber\\
&&\times(1+\frac{N_f}{6}))
\frac{165}{8}(1+\frac{5N_f}{12})(1-\frac{2N_f}{33})\ln[\frac{\mu_{MS}\beta}{2\pi}]\bigg](\frac{\alpha_s}{\pi})^2\nonumber\\
&&+(1+\frac{N_f}{6})^{\frac{1}{2}}\bigg[-799.2-21.99N_f-1.926N_f^2\nonumber\\
&&+\frac{495}{2}(1+\frac{N_f}{6})(1+\frac{2N_f}{33})\ln[\frac{\mu_{MS}\beta}{2\pi}]\bigg](\frac{\alpha_s}{\pi})^{\frac{5}{2}} \bigg \rbrace\nonumber\\
&& +O(\alpha_s)^{3}\ln (\alpha_s)). \nonumber\\
\end{eqnarray}
while EOS$\delta$ is given by
\begin{eqnarray}
\label{eqn2}
P_{g^6\ln(1/g)} &=& P_{g^5}+\frac{8\pi^2}{45}T^4 \biggl[1134.8+65.89 N_f+7.653 N_f^2\nonumber\\
   &&-\frac{1485}{2}\left(1+\frac{1}{6} N_f\right)\left(1-\frac{2}{33}N_f\right)
\ln(\frac{\mu_{MS}}{2\pi T})\biggr]\nonumber\\
&&\times\left(\frac{\alpha_s}{\pi}\right)^{3}(\ln \frac{1}{\alpha_s}+\delta).
\end{eqnarray}

As mentioned earlier, $\delta$ is an empirical  parameter, introduced to incorporate phenomenologically the
undetermined contributions at $O(g^6)$. It also acts as  a fitting parameter to get the best
agreement with the lattice results. 
\begin{figure}[htb]
\label{fig1}
\vspace*{-80mm}
\hspace*{-50mm}
\psfig{figure=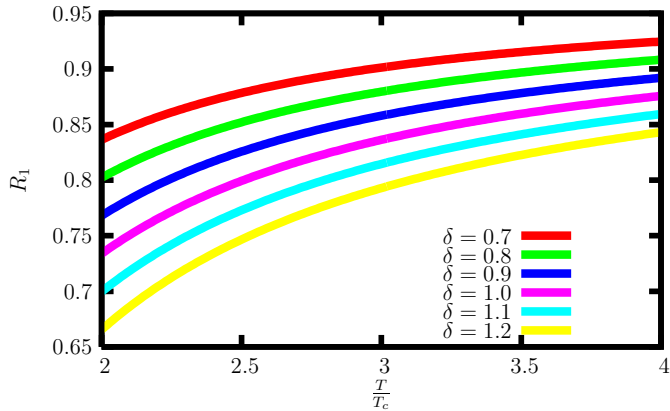,width=150mm}
\vspace*{-90mm}
\caption{(Color online) Relative equation of state({\it wrt} ideal EOS) for pure gauge theory plasma 
as a function of $T/T_c$ for various values of $\delta$.}
\end{figure}.

\begin{figure}[htb]
\label{fig2}
\vspace*{-80mm}
\hspace*{-50mm}
\psfig{figure=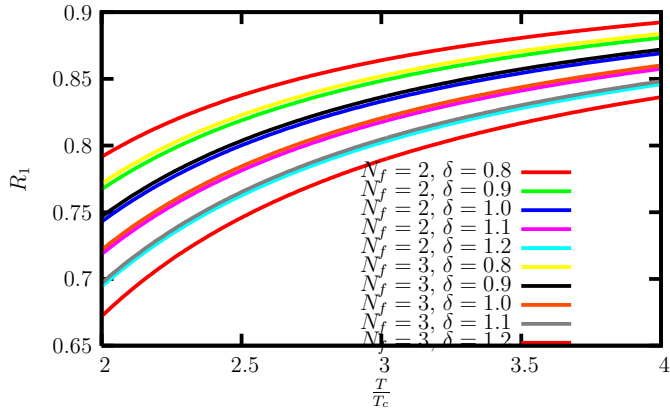,width=150mm}
\vspace*{-90mm}
\caption{(Color online) Relative equation of state{\it wrt} ideal EOS for full QCD plasma 
with $N_f=2,3$ as a function of $T/T_c$  for various values of $\delta$.} 
\end{figure}.

\subsection{The underlying distribution functions}
The  construction of  the distribution functions that underlie the
EOS, in terms of effective quarks and gluons which act as quasi-excitations, has been discussed by
Chandra et. al., \cite{chandra1} in the specific context of EOS1 and EOS$\delta$. To review the method briefly, {\sl all} the terms that represent interactions are collected together by recasting them as
effective fugacities ($z_{q,g} \equiv\exp(\mu_{q,g})$) for the otherwise free quarks and gluons. Of course, the pure gague theory
case is simply obtained by putting the number of flavors, $N_F=0$ in the EOS. Thus, $\mu_g$ represents the
self interactions of the gluons, while $\mu_f$ encapsulates the quark-quark and the quark
gluon interaction terms. Importantly, the two EOS of interest to us are valid when $T>2T_c$, and in this range, the quantities
$\tilde\mu_{q,g} \equiv \beta\mu_{q,g}$ are perturbative parameters. Thus, it is possible to solve for
$\tilde\mu_{f,g}$  self consistently through a systematic iterative procedure. In this procedure,  all the temperature
effects are contained in the effective fugacities  $z\equiv z(\alpha_s(T/T_c))$, where we display the dependence on the temperature and coupling constant  explicitly. It has been shown in Ref. \cite{chandra1} (where the details can be found) that one can trade off the dependence of the effective fugacities on the renormalization scale ( $\mu_{\bar{MS}}$) by their dependence on the  
critical temperature $T_c$. For that purpose, one utilizes the one loop expression of $\alpha_s(T)$ at finite temperature given by \cite{shaung}

\begin{eqnarray}
\label{eqn2a}
\alpha_s (T) &=& \frac{1}{8\pi b_{0}\log(T/\lambda_T)} = \alpha_s (\mu^2) \vert_{\mu=\mu_{\bar{MS}}(T)} 
\nonumber\\
\mu_{\bar MS}(T)&=& 4 \pi T \exp(-(\gamma_E+1/22)) \nonumber\\
\lambda_T &=& \frac{\exp(\gamma_E+1/22)}{4\pi}\lambda_{MS}.
\nonumber\\
\end{eqnarray}
employing which  the dependence on $\mu_{\bar{MS}}$ is eliminated, in favor of $T_c$. Consequently,
the effective chemical potentials get to depend only on  $T/T_c$.  Note that effective fugacities have merely 
been introduced to capture the interaction effects present in hot QCD equations of state. 

 Once the distribution functions are in hand, the study of transport
properties is a straight forward exercise if we employ the analysis put forth by Ranjan et. al. \cite{akranjan}. 

 In  Figs. 1 and 2,  we display the behavior of
EOS$\delta$ for various values of the parameter $\delta$. The figures show the  pure gauge theory contributions to the EOS and full QCD separately. We remark parenthetically that the studies in the earlier  work \cite{chandra1} were confined to EOS1 and the special case $\delta=0$ in EOS$\delta$. For the details on EOS1 and EOS$\delta$ for $\delta=0$, we refer the reader to Ref. \cite{chandra1} (see Fig.1-7 of Ref.\cite{chandra1}).
 First of all, we see that as $\delta$ increases in magnitude, the EOS, for both pure gauge theory and full
QCD, become softer, with $P/P_I$ taking smaller values, we denote the the ratio $P/P_I$ by $R_1$.
Kajantie\cite{kaj1} 
obtains  the best fit with the lattice results of Boyd et. al.\cite{boyd} 
by choosing a value $\delta =0.7$. We find  that to get agreement with the
more recent results of Karsch \cite{fkarsch}, $\delta \approx 1.0$ is preferred, when we consider $T >2T_c$. In short, we find that the range of values $0.8 \le \delta \le 1.2$ gives a reasonably good qualitative agreement with the lattice results for the screening lengths.

Here, we wish to mention that there is an uncertainty in fixing the free parameter $\delta$. This follows from the 
freedom in choosing the  QCD renormalization scale at high temperature. This has been investigated in detail by Blaizot, Iancu and Rebhan \cite{rebh}.
 The value of $\delta$ in the present paper has been obtained by employing the one loop expression for the running coupling constant and the QCD renormalization scale determined  in Ref.\cite{shaung}. We intend to study the quasi-particle content of 
HTL and HDL equations of state\cite{rebh,rebh1} and lattice equation of state in future.

The behavior of the corresponding fugacities, as a function of temperature, is shown in Fig.3. 
 It may be seen that $0<z_{g,q}<1.0$ which ensures the convergence of the method to determine the effective fugacities from the hot QCD EOS. We now proceed to determine the response of the plasma in the next section.

\begin{figure}[htb]
\label{fig3}
\vspace*{-80mm}
\hspace*{-50mm}
\psfig{figure=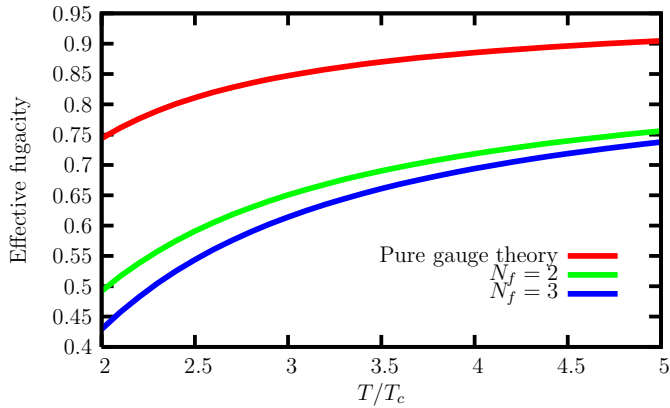,width=150mm}
\vspace*{-90mm}
\caption{(Color online) Effective parton fugacities $(z_{g,q})$ quarks determined from EOS$\delta$ as a function of temperature. Note that the behavior is shown for  $\delta=1.0$.
}
\end{figure}.

\section{Response functions for interacting qgp}
Recently Ranjan and Ravishankar \cite{akranjan} have determined the form of 
chromo-electric response functions for collision less quark-gluon plasma within 
the framework 
of semi-classical transport theory. They have set up the transport equation in 
the extended phase space including the SU(3) group space corresponding to 
dynamical color degree of freedom. They have taken the distribution function
in a coherent state basis defined over the extended single particle phase space 
$\mathcal{R}^6 \otimes \mathcal{C}_{\mathcal{G}}$, where 
$\mathcal{C}_{\mathcal{G}} = \mathcal{G} /\mathcal{H}$ is the phase space 
corresponding to the color degree of freedom, obtained as a coset space by 
factoring the group space by the stabilizer group $\mathcal{H}$ of any 
reference state in the Hilbert space. Having been employed to study the ideal 
case, the formalism has not been applied to examine the behavior of the 
plasma with a realistic EOS. We employ the results of the previous section and 
rectify this drawback, by incorporating the interaction effects as represented 
by EOS1 and EOS$\delta$.

A brief comment on the response functions. In contrast to electrodynamic plasma, 
the chromo-electric response has a richer structure. Apart from the standard 
permittivity which we shall call Abelian and denote by $\epsilon_A$, there are 
additional response functions, their number depending on the color carried by 
the partons. Thus, quarks have an additional response function which affects the 
non-Abelian coupling. The corresponding permittivity will be called non-Abelian, 
and denoted by $\epsilon_N$. The two functions exhaust the response in the quark 
sector. The gluonic sector, arising from the adjoint representation of the gauge 
group admits yet another kind of response, corresponding to tensor excitations. 
These excitations are not allowed in the quark sector (which emerges from the 
fundamental representation of the gauge group).
We consider each of these response functions for the interacting QGP. The 
response functions are obtained in the temporal gauge. 

Consider first the familiar Abelian component of the response $\epsilon_A$. For 
an isotropic plasma(in the absence of chromo-magnetic fields), its expression is 
given by \cite{akranjan}
\begin{eqnarray}
\label{eqn3}
\tilde{\epsilon}_A(\omega,\vec{k})=1+Q^2I_0(\omega,\vec{k})
\end{eqnarray}
where $Q^2 = Q^aQ^a$ is the color charge magnitude squared, and $I_0$ is 
determined by the equilibrium distribution function thus:
\[ \int \frac{1}{\omega-\frac{\vec{k}\cdot\vec{p}}{\varepsilon}}\frac{\partial f_{eq}}{\partial p_i} \,d^3\vec{p}\equiv k_iI_0(\omega,\vec{k}),\] 

The non-Abelian response function, which has been evaluated in the long 
wavelength limit, is given by
\begin{eqnarray}
\label{eqn4}
\tilde{\epsilon}_N(\omega,\omega')=\{1+\frac{Q^2\left.I_1(\omega',\vec{k}')\right|_{\vec{k}'=0}}{\omega}\}
\end{eqnarray}
where $I_1$ is defined as
\[I_1(\omega,\vec{k})=\frac{1}{3}Tr\left(\int \frac{\frac{p_j}{\varepsilon}}{(\omega-\frac{\vec{k}\cdot\vec{p}}{\varepsilon})}\frac{\partial f_{eq}}{\partial p_i} \,d^3\vec{p}\right).\]
We recall that the new constitutive Yang-Mills equations, in the presence of 
the medium, are given by

\begin{eqnarray}
\label{eqn5}
\tilde{\rho}^a(\omega,\vec{k})+iQ^2\tilde{E}^a_i(\omega,\vec{k})k_iI_o(\omega,\vec{k})\nonumber \\
-\frac{Q^2f^{alm}}{\omega} \int \left.I_1(\omega',\vec{k}')\right|_{\vec{k}'=0} \tilde{A}^l_i(\omega-\omega',\vec{k}-\vec{k}')\nonumber \\
\times \tilde{E}^m_i(\omega',\vec{k}') \,d\omega' \,d^3\vec{k}'=0.
\end{eqnarray}

\begin{eqnarray}
\label{eqn6}
\tilde{j}^a_j(\omega,\vec{k})+iQ^2\tilde{E}^a_i(\omega,\vec{k})\delta_{ij}\left.I_1(\omega,\vec{k})\right|_{\vec{k}=0}\nonumber \\
=0.
\end{eqnarray}

As pointed out in \cite{akranjan}, the Abelian and non-Abelian responses are not 
independent of each other. Gauge invariance relates them, by virtue of which we 
can obtain both from a common generating function as follows:

\begin{eqnarray}
\label{eqn7}
I_0= \frac{1}{k^2}\frac{\partial}{\partial \omega} \int ln(\omega -
\frac{\vec{k}\cdot \vec{p}}{\varepsilon})k_i \partial_{{p_i}}f_{eq} d^3p \nonumber \\
I_1= -\frac{1}{3}Tr\bigg(\frac{\partial}{\partial k_j} \int ln(\omega -
\frac{\vec{k}\cdot \vec{p}}{\varepsilon}) \partial_{p_i}f_{eq} d^3p\bigg).
\end{eqnarray}

We further recall that these expansions are determined when the system is 
displaced slightly from its equilibrium, in the collisionless limit.

\subsection{Ideal response}
It is convenient to first write the expressions for the responses of ideal 
distributions for quarks and gluons. The responses due to EOS1 and EOS$\delta$ 
get a simple modification over their ideal forms since we have mapped 
successfully the interaction effects into quasi free partons with effective 
fugacities. Thus, in the ideal case we have, for the quarks,

\begin{eqnarray}
\label{eqn8}
\tilde{\epsilon}_A^{(q)}=[1+\frac{2\pi^3Q^2T^2N_f}{3k^2}\{-\frac{\omega}{k}ln\left|\frac{\omega+k}{\omega-k}\right|+2\}]\nonumber\\
\end{eqnarray}

and the non-Abelian response function is given by
\begin{eqnarray}
\label{eqn9}
\tilde{\epsilon}_N^{(q)}
=\{1-\frac{4\pi^3 Q^2T^2N_f}{9}\frac{1}{\omega\omega'}\}.
\end{eqnarray}


The imaginary part of Abelian($\tilde\epsilon_A$) and non-Abelian component ($\tilde\epsilon_N$) of the chromo-electric permittivity can be easily evaluated by the standard Landau $i\epsilon$ prescription. These are needed to obtain landau damping which we do not study here.

The contribution to the permittivity from  the gluons is closely related, and not independent of the contribution of the quarks written above. Indeed, if we define the susceptibilities
$${\cal A}^{(q,g)} = \tilde{\epsilon}_A^{(q,g)}-1$$
and 
$${\cal N}^{(q,g)} = \tilde{\epsilon}_N^{(q,g)}-1$$
for the quarks and the gluons,  It can be shown that \cite{akranjan} the gluonic permittivity can be simply
read off from the quark permittivity (and vice versa) as
\begin{eqnarray}
\label{susdef}
{\cal A}(q)=\frac{N_f}{2}{\cal A}(g),~~
{\cal N}(q)=\frac{N_f}{2}{\cal N}(g).
\end{eqnarray}
where $N_F$ is the number of flavors.
In short, for the total susceptibility, we have the simple relation
 $
\chi^{A,N}_q =\frac{N_F}{2}\chi^{A,N}_g,
$

\subsection{Interaction effects}
We now consider the modification that the above expressions undergo permittivities arising because of the new EOS. Recall 
that the corresponding equilibrium distribution functions differ from each 
other only in their form for the chemical potentials $\mu_{q,g}$. The responses 
thus depend on the interactions implicitly through an explicit dependence 
on $z_{q,g}$.

 Considering the gluonic case, {\sl i. e.}, pure gauge theory first, we 
get the expressions for the two permittivities as

\begin{eqnarray}
\label{eqn10}
\tilde{\epsilon}_A=[1+\frac{2\pi^3 Q^2T^2 g_2^{\prime}(z_g)}{3k^2}\{-\frac{\omega}{k}ln\left|\frac{\omega+k}{\omega-k}\right|+2\}],
\end{eqnarray}

 and the non-Abelian response function is 

\begin{eqnarray}
\label{eqn11}
\tilde{\epsilon}_N
=\{1-\frac{4\pi^3 Q^2T^2 g_2^{\prime}(z_g)}{9}\frac{1}{\omega\omega'}\}.
\end{eqnarray}
The function  $g_2^{\prime}(z_g) \equiv \frac{6}{\pi^2} g_2(z_g)$ where 
$g_2(z_g)$ is defined via the integral  below.

$$
\int_0^{\infty} \frac{x^{\nu-1}}{z_g^{-1}exp({x})-1}\,dx=\Gamma(\nu)g_{\nu}(z_g)\
$$
$g_{\nu}(z_g)$ has the series expansion
$$
g_{\nu}(z_g)=\sum_{l=1}^{\infty} \frac{z_g^l}{l^{\nu}} ~~\mbox{for}~~ z_g\ll 1.
$$
Note that $g'_2(1)=1$ gives the ideal limit.

Similarly, the corresponding expressions for in the quark sector are obtained as

\begin{eqnarray}
\label{eqn12}
\tilde{\epsilon}_A=[1+\frac{2\pi^3Q^2T^2N_f f^{\prime}_2(z_f)}{3k^2}\{-\frac{\omega}{k}ln\left|\frac{\omega+k}{\omega-k}\right|+2\}]\nonumber\\
\end{eqnarray}

and the non-Abelian response for effective quarks reads:

\begin{eqnarray}
\label{eqn13}
\tilde{\epsilon}_N
=\{1-\frac{4\pi^3 Q^2T^2N_f f^{\prime}_2(z_f)}{9}\frac{1}{\omega\omega'}\}.
\end{eqnarray}

The function  $f_2^{\prime}(z_f) \equiv \frac{12}{\pi^2} f_2(z_f)$ where 
$f_2(z_f)$ is defined via the integral below.

\[\int_0^{\infty} \frac{x^{\nu-1}}{z_f^{-1}exp({x})+1}\,dx=\Gamma(\nu)f_{\nu}(z_f)\]
\[f_{\nu}(z_f)=\sum_{l=1}^{\infty} (-1)^{l-1} \frac{z_f^l}{l^{\nu}} ~~\mbox{for}~~ z_f\ll 1\]
and $f'_2(1)=1$.
 
\section{Effective charges and relative susceptibilities}
Eq.(\ref{eqn10}-\ref{eqn13}) admit a simple physical interpretation, when compared with their
counterparts Eq.(\ref{eqn8} -\ref{eqn9}). Indeed, the sole effect of the interactions on the transport properties is to merely renormalize the   
the quark and the gluon charges $Q_{g,q}$ as shown 
below:
$$
Q_g^2 \rightarrow \bar{Q}_g^2 = Q^2 g_2^{\prime}(z_g); ~~Q_q^2 \rightarrow \bar{Q}_q^2 = Q_q^2 f_2^{\prime}(z_f).
$$
The renormalization factors $g_2^{\prime}(z_g), f_2^{\prime}(z_f)$ further possess the significance of
chromo-electric susceptibilities, relative to the ideal values. To see that, we 
note that the Abelian and the non-Abelian strengths for gluons as well as quarks 
suffer the same renormalization reflecting the underlying gauge invariance. 
Furthermore, the expressions for the relative susceptibilities are given by, 
\begin{eqnarray}
\label{eqn14}
{\cal R}=\frac{\chi(z)}{\chi(1)}\equiv \frac{{\cal A}(z)}{{\cal A}(1)}=\frac{{\cal N}(z)}{{\cal N}(1)}= \left\{ \begin{array}{rcl}
f'_2(z_f) &\mbox{for quarks,} &\\
g'_2(z_g) &\mbox{for gluons}&
\end{array} \right.
\end{eqnarray}
and
\begin{eqnarray}
\label{eqn15}
{\cal R}_{q,g}=\frac{\chi^{(q)}(z_f)}{\chi^{(g)}(z_g)}\equiv\frac{{\cal A}^{(q)}(z_f)}{{\cal A}^{(g)}(z_g)}=\frac{{\cal N}^{(q)}(z_f)}{{\cal N}^{(g)}(z_g)}=\frac{f'_2(z_f)N_f}{g'_2(z_g)}.
\end{eqnarray}
Note that the relative susceptibilities are entirely functions of the single variable $T/T_c$, and are
independent of ($\omega, k$). The dependence of the susceptibilities on ($\omega, k$) has already been studied in
detail in Ref.\cite{akranjan}. We merely concentrate on the temperature dependence below.

Before we go on to discuss the susceptibilities and other bulk properties, 
 we point out an essential care to be taken in using the above 
susceptibilities for determining the response of the plasma. For pure gauge theory, only 
the gluonic part contributes, while for the full QCD, we have to necessarily 
take the contribution from both the quark and the gluonic sector. We discuss 
both the cases below. The response functions for the full QCD is obtained by 
averaging up the above calculated response functions for quark as well as 
gluon plasma. The relative susceptibility for full QCD plasma is given by
\begin{eqnarray}
\label{eqn16}
{\cal R'}=\frac{\chi(\tilde{z})}{\chi(1)}\equiv \frac{{\cal A}(\tilde{z})}{{\cal A}(1)}=\frac{{\cal N}(\tilde{z})}{{\cal N}(1)}\nonumber\\
=\frac{N_ff'_2(z_f)+2g'_2(z_g)}{N_f+2},
\end{eqnarray} 
where $\tilde{z}$ is the effective fugacity of partons in full QCD plasma.

\subsection{Behavior of the susceptibilities}
We now proceed to study the behavior of the relative susceptibilities displayed 
in Eqs.(\ref{eqn14}), (\ref{eqn15}) and (\ref{eqn16}) as  functions of temperature.  As observed, relative susceptibilities for both quarks and 
gluons scale with $T/T_c$. 
We have plotted the relative susceptibilities ${\mathcal R}$, 
${\mathcal R_{qg}}$ and ${\mathcal R}^\prime$ as  functions of $T/T_c$
(See Figs.4-7), for both EOS1 and EOS$\delta$.  Please note that 
we have chosen $\delta=1.0$ in EOS$\delta$.

Fig.4 shows the relative susceptibility of a purely gluonic plasma as a function of 
temperature for EOS1 and EOS$\delta$.

\begin{figure}[htb]
\label{fig4}
\vspace*{-70mm}
\hspace*{-45mm}
\psfig{figure=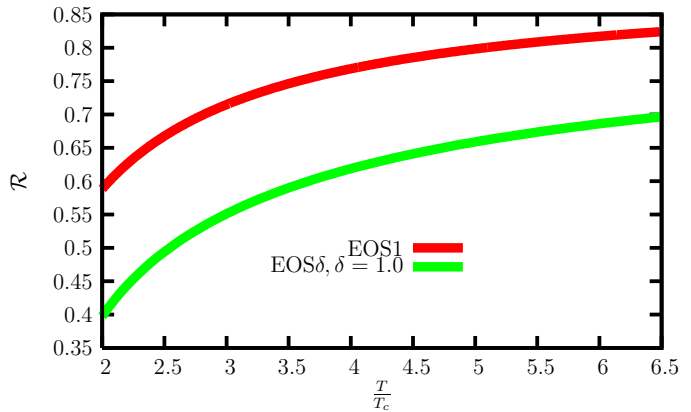,width=150mm}
\vspace*{-80mm}
\caption{(Color online) Relative susceptibility, $g'_2(z_g)$ (see Eq. (\ref{eqn14}), for  pure gauge theory plasma 
 as a function of $T/T_c$ for EOS1 and EOS$\delta$ ($\delta=1$).}.  
\end{figure}

\begin{figure}[htb]
\label{fig6}
\vspace*{-80mm}
\hspace*{-50mm}
\psfig{figure=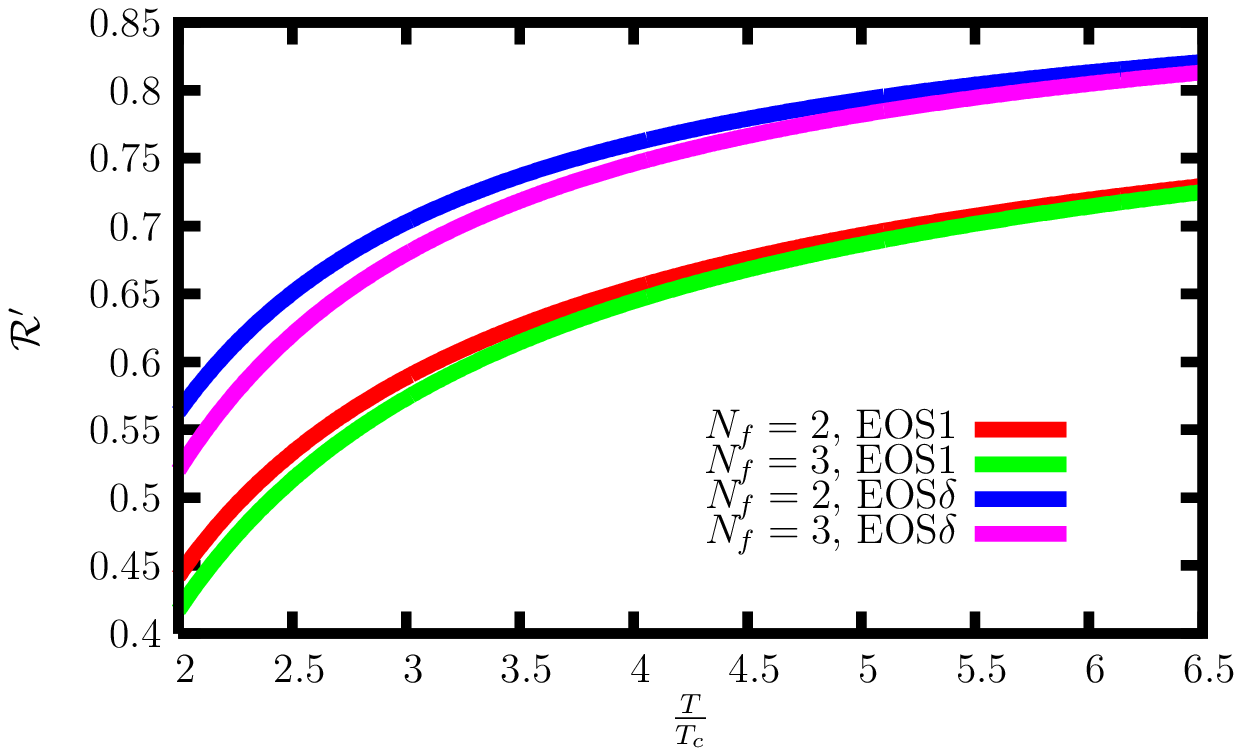,width=150mm}
\vspace*{-90mm}
\caption{(Color online) Relative susceptibility, defined in Eq. (\ref{eqn16}), for the full QCD plasma 
 as a function of $T/T_c$, for EOS1 and EOS$\delta$ ($\delta=1)$. We have studied the cases $N_f= 2,3$.}
\end{figure}  
We see From Fig. 4 that the susceptibility of a purely gluonic plasma is weaker in the presence of interactions, approaching its ideal value asymptotically
with increasing temperatures. Equivalently, there is a decrease in the value of the phenomenological coupling
$Q^2$, relative to its ideal value.

The behavior of quark gluon plasma is not qualitatively different from that of a purely gluonic plasma,
as may be seen from Fig.5. In other words, the quark contribution is of the same order as the purely gluonic contribution. However, the relative contribution from the quarks and the gluons does depend on the EOS considered. Indeed, with EOS1 (where interactions up to $O(g^5)$ are included), Fig.6 shows that the quark contribution
dominates slightly over the gluonic  contribution for $N_F=2$. The dominance is more pronounced for the more realistic case $N_F=3$. 
In contrast, we see from Fig. 7, that EOS$\delta$ (with $\delta=1$) predicts that the gluonic contribution is marginally larger for $N_F=2$ and becomes sub dominant when $N_F=3$. This distinction between the two
EOS is of no practical consequence since, given $T_c \sim 170 MeV$, one has to necessarily work with 
$N_F=3$ at $T =2T_c$.
\begin{figure}[htb]
\label{fig5}
\vspace*{-80mm}
\hspace*{-45mm}
\psfig{figure=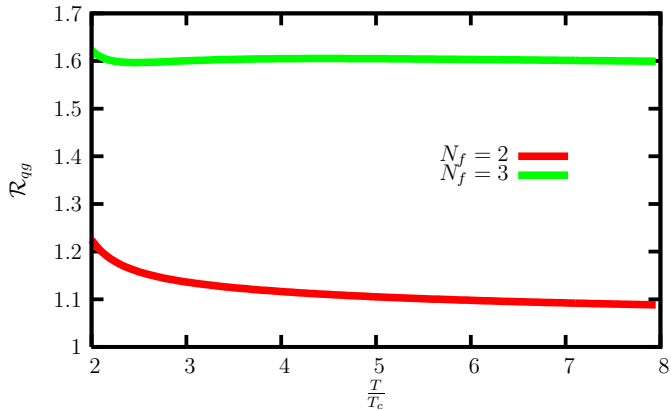,width=150mm}
\vspace*{-90mm}
\caption{(Color online) Ratio of the quark to gluonic contributions to the susceptibility (see Eq.(\ref{eqn15}) 
as a function of $T/T_c$, as predicted by EOS1.
}
\end{figure}.

\begin{figure}[htb]
\label{fig7}
\vspace*{-70mm}
\hspace*{-50mm}
\psfig{figure=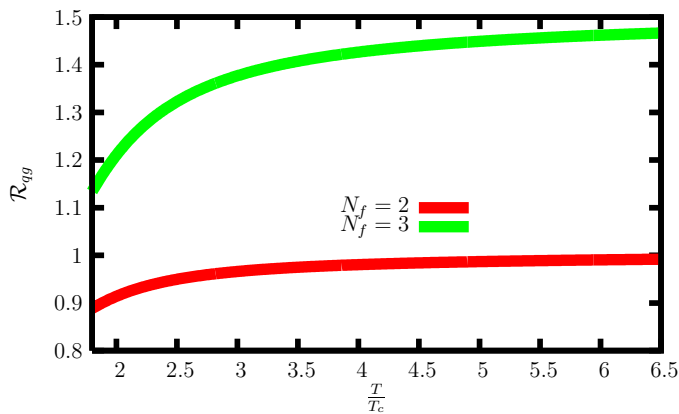,width=150mm}
\vspace*{-90mm}
\caption{(Color online) Ratio of the quark to gluonic contributions to the susceptibility (see Eq.(\ref{eqn15}) 
as a function of $T/T_c$, for EOS$\delta$, $\delta=1$..}
\end{figure}

\section{The heavy quark potential}
Now we shall apply the results of the previous sections to discuss the heavy 
quark potential in the presence of interacting  medium. We consider the 
Cornell potential
 \[\phi(r)=-\frac{\alpha}{r}+\Lambda r \]
where $\alpha$ and $\Lambda$ are  phenomenological constants. The first term
shows the Coulombic behavior and dominates at small distance while
the second term causes  linear confinement, dominating at large distances.

It had been expected earlier that the long range part of the Cornell potential
does not survive in the quark gluon phase. This expectation assumes  a phase transition from the 
hadronic to deconfined phase. More recent studies\cite{phaseT} indicate that in all likelihood,  deconfinement is not a phase transition, but a crossover. If such to be the case, there is no reason to expect the 
linear part of the potential to disappear completely. With this in mind, we study the 
modifications of both the Coulomb and linear terms, and examine how reasonable the EOS under
consideration are.



Since the potential has no explicit color dependence, it is sufficient to employ the Abelian components
of the permittivities. At $\omega=0$, the quark and gluon permittivities have the form
\begin{eqnarray}
\label{eqn16a}
\tilde\epsilon_q (k,T)=1+\frac{16\pi{Q^2}T^2}{k^2}f_2(z_q)\nonumber\\
\tilde\epsilon_g(k,T)=1+\frac{16\pi{Q^2}T^2}{k^2}g_2(z_g).\nonumber\\
\end{eqnarray}

Therefore the full permittivity  reads
\begin{eqnarray}
\tilde\epsilon(k,T)&=&\frac{(\tilde\epsilon_g +\tilde\epsilon_q)}{2} \nonumber\\
&=&1+\frac{8\pi^2{Q^2}T^2}{k^2}\bigg[N_f f_2(z_q)+g_2(z_g)\bigg] \nonumber \\ &\equiv & 1 + \frac{m_D^2}{k^2},
\end{eqnarray}
in terms of the Debye mass $m_D^2 = 8\pi Q^2 T^2\big[N_f f_2(z_q) +g_2(z_g)\big]$.

The $q\bar{q}$ potential undergoes a modification due to the medium via
$\tilde\epsilon(k,T)$, as given by  $\tilde{\phi}(k) \rightarrow \tilde{\phi}(k)/\tilde{\epsilon}(k,T)
\equiv \tilde{\phi}_s(k,T)$.  We note that in determining the 
 Fourier transform of Cornell potential, we regulate  the 
linear term exactly the same way we regulate the Coulomb term, by multiplying with an exponential
damping factor.   The damping is 
switched off after the Fourier transform is evaluated. The Fourier transform 
is thus obtained as  
\begin{eqnarray}
\label{eqn16b}
\tilde{\phi}(k)=-\sqrt{\frac{2}{\pi}}\frac{1}{k^2}-\frac{4\Lambda}{k^4\sqrt{2\pi}}
\end{eqnarray}

The modified potential thus acquires the form
\begin{eqnarray}
\label{eqn17}
\tilde{\phi}_s(k,T)=-\sqrt{\frac{2}{\pi}}\frac{\alpha}{k^2+m^2_D}
-\frac{4}{\sqrt{2\pi}}\frac{\Lambda }{k^2(k^2+m^2_D)}.
\end{eqnarray}
We note that for a gluonic plasma,
$m^2_{D}=16\pi Q^2 T^2 g_{2}(z_g)$.

On comparing Eq.\ref{eqn17} with Eq.\ref{eqn16b} we infer the renormalization of the couplings  
$$
\Lambda_{eff}=\frac{\Lambda}{1+\frac{m^2_D}{k^2}};~~
\alpha_{eff}=\frac{\alpha}{1+\frac{m^2_D}{k^2}}.
$$

\subsection{Screening of the heavy quark potential}
Of interest to us is the form of the potential in the real space, as a function of
spatial separation. The inverse Fourier transform yields it to be
\begin{eqnarray}
\label{eqn21}
\phi_s(r,T)&=&(\frac{2\Lambda}{m^2_D}-\alpha)\frac{\exp{(-m_Dr)}}{r}\nonumber\\
&-&\frac{2\Lambda}{m^2_Dr}+\frac{2\Lambda}{m_D}-\alpha m_D.
\end{eqnarray}
It follows from the above equation that the medium transforms the linear potential to the long
range Coulomb form, just as it modifies the bare Coulomb term to the short ranged Yukawa..
The modified potential is not short ranged; it is not confining either.
To appreciate  this, note that at large $T$, the above expression reduces to
\begin{eqnarray}
\label{lrp}
\phi_s(r,T)\sim -\frac{2\Lambda}{m^2_Dr}-\alpha m_D
\end{eqnarray}
 Thus, contrary to the Maxwellian
plasmas which support only short range interactions, the two EOS predict that the heavy quark potential continues to be long ranged,
although  absolute confinement, which was a quintessential feature of the unscreened potential, is lost.
It might as well be that the above results signify that the EOS fail to describe the hadronic matter in its
deconfined state, On the other hand, since the transition from the hadronic to QGP phase could be a cross over, and not a phase transition \cite{phaseT}, it could be possible that the above result is not entirely devoid
of physical significance.  If we adopt the latter view, if only for the purposes of analysis,
a discussion of screening cannot, therefore rely entirely on the interpretation of inverse Debye mass as a screening length
in its usual sense. We address the issue below .  

Let us consider the high temperature limit of the potential, given by Eq.(\ref{lrp}).
Ignoring the additive contribution, the  energy of the $q\bar{q}$ in the ground state is simply given
by
\[E_g=\frac{m_q\Lambda^2}{m^4_D},\]
where $m_q$ is the mass of heavy quark.
\begin{figure}[htb]
\label{fig8}
\vspace*{-75mm}
\hspace*{-45mm}
\psfig{figure=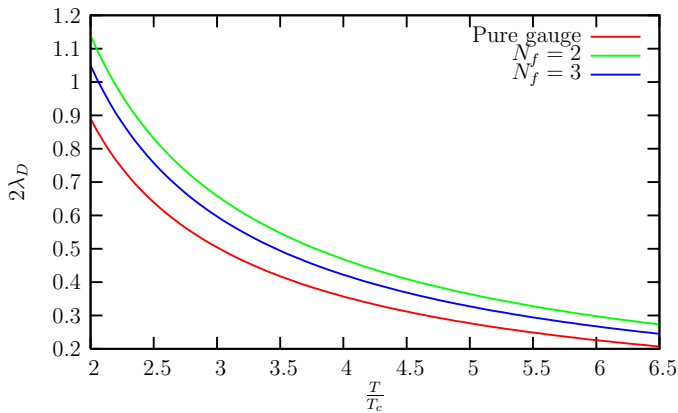,width=150mm}
\vspace*{-90mm}
\caption{ (Color online) Debye screening length for gluonic and quark-gluon plasmas as a function of 
$T/T_c$ for EOS1. }
\end{figure}

\begin{figure}[htb]
\label{fig9}
\vspace*{-75mm}
\hspace*{-42mm}
\psfig{figure=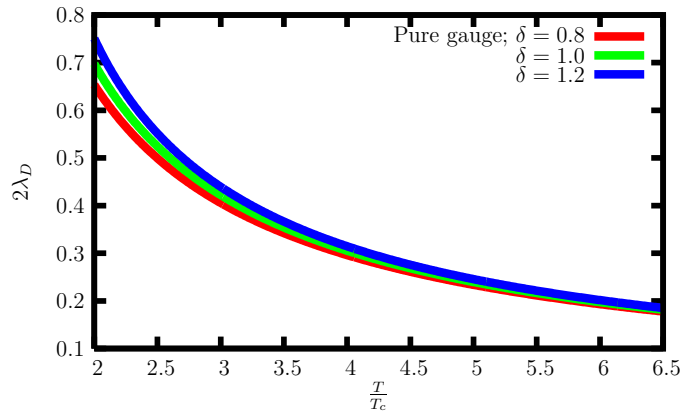,width=150mm}
\vspace*{-90mm}
\caption{(Color online) Debye screening length for pure gauge theory plasma in EOS$\delta$  as a function of 
$T/T_c$ for various values of $\delta$.}
\end{figure}
The binding energy is, of course, temperature dependent and approaches zero as $T \rightarrow \infty$.
At any finite temperature though, the quarks possess a thermal energy $E_T \sim T$ (by equipartition theorem),
leading to an ionization of the quarkonium when $E_T$ matches the binding energy. The dissociation temperature
$T_d$ is determined by the matching conditions. In the case of pure gauge theory, 
\begin{equation}
\label{eqn22}
\frac{m_q \Lambda^2}{384\pi^2Q^4T^5_c}= (\frac{T_D}{T_c})^5 g^2_2(z_g).
\end{equation}

And for full QCD:
\begin{equation}
\label{eqn23}
\frac{m_q \Lambda^2}{384\pi^2Q^4T^5_c}=\frac{1}{4}(\frac{T_D}{T_c})^5 \bigg(N_ff_2(z_q)+g_2(z_g)\bigg)^2 
\end{equation}

\subsection{Estimation of $Q$ and a determination of the screening length}
The above equation is still not amenable to comparison with experiments since it has the undetermined parameter $Q$. To estimate $Q$, we need an additional input which we obtain by comparing the screening length
obtained as a solution of Eqs. (\ref{eqn22}) and (\ref{eqn23}) with the lattice results, reported 
by Kazmarek and Zantow \cite{zantow}.  Note that the screening lengths  for  gluonic and quark gluonic plasmas
are respectively given by
\begin{eqnarray}
\label{eqn24}
\lambda^g_D=\frac{1}{QT}\frac{1}{\sqrt\bigg(16\pi g_2(z_g)\bigg)},
\end{eqnarray}
\begin{eqnarray}
\label{eqn25}
\lambda_D=\frac{1}{ QT}\frac{1}{\sqrt\bigg(8\pi(N_f f_2(z_q)+g_2(z_g))\bigg)}.
\end{eqnarray}

Recalling that our results are valid for $T > 2T_c$, we match the pure gauge theory result with the
lattice values $\lambda \sim 0.15$ fm, and $T_c = 0.27$ GeV. We obtain
\begin{eqnarray}
\label{eqn26}
\lambda^g_D=\frac{0.19}{0.27}\frac{1}{4Q(T/T_c)}\frac{1}{\sqrt(\pi g_2(z_g))}.
\end{eqnarray}
This leads to the estimate $Q \sim 0.15$. The temperature dependence of the screening lengths can
be thereafter determined for the two equations of state. We emphasize that the choice $\delta =0.9$
gives the best agreement between EOS$\delta$ and the lattice EOS for gluonic plasma \cite{fkarsch}.

\subsection{Dissociation temperatures for quarkonia}
Since there are no free parameters left, it is a straight forward task to determine the dissociation temperatures for the heavy quark bound states. We are principally interested in $J/\Psi$ and $b\bar{b}$
states, for which we have gathered the results in Table. 1, after obtaining graphical solutions
for Eq.(\ref{eqn22}) and Eq.(\ref{eqn23}). We have employed the values $m_c = 1.5 GeV,~m_b=4.5 GeV$ and $\Lambda = 0.18 GeV^2$
for the quark masses and the strength of the Cornell potential. It is noteworthy that the dissociation temperatures are all roughly in the range $ T_D \approx (2-3)T_c$, which is higher than the temperatures
achieved so far. Since the temperatures expected at LHC is in the range $T \sim 2T_c-3T_c$, one may expect to test these predictions there. 

We now  turn our attention to compare hot QCD estimates for dissociation temperatures
with other theoretical works. In a recent paper, Satz\cite{satz} has studied the dissociation of quarkonia states by studying their in-medium behavior. These estimates were based on the Schr\"odinger equation for Cornell potential. In a more recent work,  Alberico {\it et al}\cite{prd75} reported the dissociation temperatures for charmonium and bottomonium states for $N_f=0$ and $N_f=2$
QCD. In this work, they have solved the  Schr\"odinger equation for the charmonium and bottomonium states at finite temperature in the presence of temperature dependent potential-- computed from the lattice QCD.  We have  quoted these results in Table 2.
The estimates for $N_f=0$ and $N_f=2$ cases for both EOS1 and EOS$\delta$(Table 1) are closer to Ref.\cite{satz}. On the other hand the estimates for $J/\Psi$ dissociation temperatures for both EOS1 and EOS2 are larger than that of Ref.\cite{prd75} while bottomonium dissociation temperature estimates are slightly smaller. We do not have lattice estimates at present to compare the dissociation temperatures for $N_f=3$ QCD. However the hot QCD estimates are consistent with the lattice predictions\cite{sdata} on the survival of heavy quarkonia states near $2T_c$ and  predictions  of dynamical $N_f=2$ QCD by Aarts {\it et al}\cite{gert}.
Along these results, we wish to mention the very recent estimates on dissociation temperature reported by M\'ocsy and P\'etreeczky\cite{moscky}. Their estimates for $J/\Psi$ dissociation temperature is $1.2 T_c$ and for $\Upsilon$ is $2 T_c$. The 
estimates for both EOS1 and EOS$\delta$ are larger as compare to these results.

\begin{table}
\label{table1}
\caption{The dissociation temperature($T_D$) for various quarkonia states (in unit of $T_c$).}
\centering
\begin{tabular}{|l|l|l|l|l|}
\hline
Hot EOS& Quarkonium &Pure QCD& $N_f=2$&$N_f=3$\\
&&&&\\
\hline\hline
EOS1&$J/\Psi$&2.2&2.62&2.46\\
&$\Upsilon$&2.5&3.14&2.94\\
\hline
EOS$\delta$&$J/\Psi$&1.86&2.38&2.24\\
$\delta=0.8$&$\Upsilon$&2.12&2.76&2.58\\
\hline
EOS$\delta$&$J/\Psi$&1.95&2.45&2.32\\
$\delta=1.0$&$\Upsilon$&2.2&2.83&2.66\\
\hline
EOS$\delta$&$J/\Psi$&2.03&2.52&2.40\\
$\delta=1.2$&$\Upsilon$&2.28&2.9&2.74\\
\hline
\end{tabular}
\end{table}

\begin{table}
\label{table2}
\caption{The dissociation temperature($T_D$) for various quarkonia states (in unit of $T_c$) from Ref.\cite{satz} and Ref.\cite{prd75}. The first and third rows are the estimated values for dissociation temperature from Ref.\cite{satz} and second and fourth are from Ref.\cite{prd75}}
\centering
\begin{tabular}{|l|l|l|}
\hline
Quarkonium &$N_f$=0& $N_f=2$\\
&&\\
\hline\hline
$J/\Psi$&2.1&$>$2\\
\hline
$J/\Psi$&1.40&1.45\\
\hline
$\Upsilon$&$>$2&\\
\hline
$\Upsilon$&2.96&3.9\\
\hline
\end{tabular}
\end{table}

\subsubsection{Comparison of Debye screening length with lattice results}
Finally, with a view to benchmark  our estimates of the screening lengths, by comparing them with the recent lattice results reported by Kazmarek and Zantow \cite{zantow}, we
 plot $2 \lambda_D$ as a function of $T/T_c$. The results are shown for EOS1 as well as EOS$\delta$, for three values $\delta=0.8,
1.0,1.2$. The results for pure gauge theory (gluonic plasma) are shown in Fig.9, and Fig.10 shows the results for
full QCD. We find that on comparison with Fig.2 of Ref. \cite{zantow}, these values, $\delta \sim 1$
are the most favored which justifies, {\sl a posteriori} our choice for the parameter.

\begin{figure}[htb]
\label{fig10}
\vspace*{-75mm}
\hspace*{-42mm}
\psfig{figure=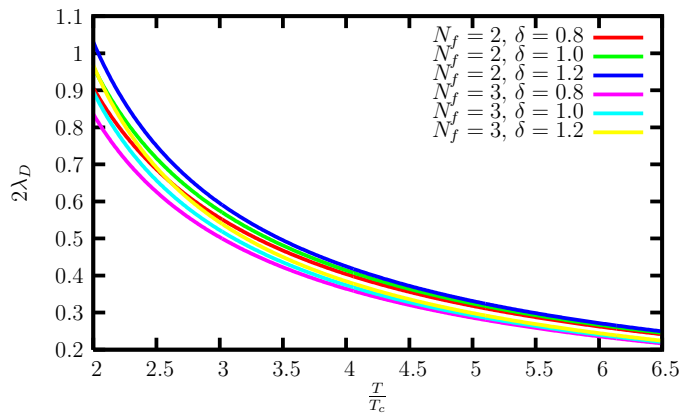,width=150mm}
\vspace*{-90mm}
\caption{(Color online) Debye screening length for full QCD plasma as predicted by EOS$\delta$ as a function of 
$T/T_c$ for various of $\delta$.}
\end{figure}
\section{Conclusions and Outlook}
In conclusion, we have successfully extracted the quasi-free particle content of two hot QCD equations of states and used them  to determine the
chromo-electric permittivities within the standard Boltzmann-Vlasov kinetic approach. The Abelian and the
non-Abelian components of the permittivity are obtained, for pure gauge theory and the full QCD. 
We have shown that the effect of the interactions is to merely renormalize the magnitude of the effective color charge, $Q$. We have used the permittivities to study critically the modifications in a realistic
heavy quark potential. The dissociation temperatures are carefully estimated, by fixing the magnitude of $Q$
by an explicit matching with a lattice result. The values obtained are quite close to the exact lattice results. The viability of the two EOS, especially EOS$\delta$ is thus phenomenologically well supported.
Our analysis suggests strongly, and in agreement with the lattice results, that $J/\Psi$ suppression can
be seen in QGP only for $T \ge 3T_c$.

A true test of the above predictions would be possible if we succeed in extracting a quasi particle description from the lattice EOS. Studies are under way in this direction.
It should also be of interest to extend the analysis to other signatures like strangeness enhancement, and also
for QGP with a finite baryonic chemical potential \cite{avrn,ipp}. These will be taken up in a later work.

\vspace{6mm}
\noindent {\bf Acknowledgments:}
 VC acknowledges Anton Rebhan for useful comments and suggestions on the work in the present manuscript
 during Les Houches QCD school-- {\bf Hadronic collisions at the LHC and QCD at high density} at Centre de Physique des Houches, France - Mar 25 - Apr 4, 2008. VC also acknowledges C.S.I.R., New Delhi (India) for the financial support.


\end{document}